# High In-content InGaN layers synthesized by plasma-assisted molecular-beam epitaxy: growth conditions, strain relaxation and In incorporation kinetics

S. Valdueza-Felip[1,2]\*, E. Bellet-Amalric[1,2], A. Núñez-Cascajero[1,2,3], Y. Wang[4],

M.-P. Chauvat[4], P. Ruterana[4], S. Pouget[1,2], K. Lorenz[5], E. Alves[5], and E. Monroy[1,2]

[1]*Université Grenoble Alpes, 38000 Grenoble, France*
[2]*CEA-Grenoble, INAC/SP2M, 17 rue des Martyrs, 38054 Grenoble, France*
[3]*GRIFO, Dept. Electrónica, Universidad de Alcalá, 28871 Alcalá de Henares, Madrid, Spain*
[4]*CIMAP, CNRS-ENSICAEN-CEA-UCBN, 6 Blvd. Maréchal Juin, 14050 Caen, France*
[5]*IPFN, Instituto Superior Técnico (IST), Campus Tecnológico e Nuclear, Estrada Nacional 10, P-2695-066 Bobadela LRS, Portugal*
\*sirona.valdueza-felip@cea.fr

We report the interplay between In incorporation and strain relaxation kinetics in high-In-content $In_xGa_{1-x}N$ ($x = 0.3$) layers grown by plasma-assisted molecular-beam epitaxy. For In mole fractions $x = 0.13$–$0.48$, best structural and morphological quality is obtained under In excess conditions, at In accumulation limit, and at a growth temperature where InGaN decomposition is active. Under such conditions, *in situ* and *ex situ* analysis of the evolution of the crystalline structure with the layer thickness points to an onset of misfit relaxation after the growth of 40 nm, and a gradual relaxation during more than 200 nm, which results in an inhomogeneous strain distribution along the growth axis. This process is associated with a compositional pulling effect, *i.e.* indium incorporation is partially inhibited in presence of compressive strain, resulting in a compositional gradient with increasing In mole fraction towards the surface.

## I. INTRODUCTION

InGaN ternary alloys, with a direct band gap tunable from 0.7 to 3.4 eV (1.7−0.36 μm), are particularly relevant for the fabrication of violet/blue/green light emitters, and intense research is underway to extend its application towards longer wavelengths. Furthermore, InGaN has been recently proposed as a potential solution for a double-junction tandem solar cell based on III-V materials on silicon [1,2], with a predicted efficiency limit of 39% under 1 sun, and 42.3% under solar concentration (500 suns). An important advantage for device processing is that III-nitrides are similar to silicon in terms of thermal budget, which renders feasible the development of a joint InGaN-on-silicon technology. However, the fabrication of such photovoltaic devices requires a high-quality InGaN junction with an In mole fraction around 30%.

The synthesis of high-quality high-In-content InGaN films is challenging due to the lattice mismatch between InN and GaN (~11%), the composition instabilities over a wide range of indium content [3], the relatively high vapor pressure of InN as compared to GaN leading to low indium incorporation [4], and the difference in formation enthalpies for InN and GaN which causes a strong indium segregation on the growth front [5]. Plasma-assisted molecular beam epitaxy (PAMBE) allows the synthesis of InGaN in the entire alloy composition range, since this growth process is far from thermodynamic equilibrium. In PAMBE, the incorporation of active nitrogen at the growth front is independent of the substrate temperature, which allows lower growth temperature than techniques requiring ammonia cracking. PAMBE growth of InGaN has been reported following various approaches. Gacevic *et al.* [6] presented a growth diagram of InGaN with In content up to 50% and identified an intermediate metal rich regime, with 1−2 monolayer (ML) thick In coverage, which provided best InGaN quality in terms of surface morphology. On the other hand, Yamaguchi *et al.* [7] showed the growth of thick and uniform $In_xGa_{1-x}N$ films with the entire alloy composition ($0.2 \leq x \leq 1$) using the droplet elimination radical-beam irradiation method. In this method, the impinging Ga and N fluxes remain constant, and the In flux is modulated, alternating growth under In excess and interruptions of the In flux to consume the excess accumulated on the surface. High In-content $In_xGa_{1-x}N$ ($0.2 \leq x \leq 0.67$) layers on GaN were also reported using metal-modulated PAMBE, growing at low substrate temperature (400−450°C), constant N flux, and modulating Ga and In fluxes under metal-rich conditions [8]. This growth method provides a significant improvement of the crystalline quality for In compositions above 60%, thanks to a full relaxation of the misfit strain through a periodic network of misfit dislocations at the GaN/InGaN interface [9].

Given the huge lattice mismatch between GaN and InN, the accommodation of misfit strain during the InGaN epilayer growth is of crucial importance to determine the final structural quality and alloy composition. A variation of the composition and strain as a function of depth has been observed in $In_xGa_{1-x}N$ ($x \sim 0.2$) layers [10−12] and InGaN/GaN multiple-quantum wells [13] grown by metal-organic vapor phase epitaxy, as well as InGaN structures grown by PAMBE [14]. In all cases, the In mole fraction increases for thicker samples.

In this work, we address the interplay between In incorporation and strain relaxation in high-In-content InGaN layers grown by PAMBE on GaN. In particular, we analyze InGaN films with an In mole fraction around 30%, the composition required to complement the response of the silicon solar cell in a double-junction system [2]. The effect of the strain and composition distribution on the optical properties of the layers is also presented.

## II. EXPERIMENTAL METHODS

InGaN samples were grown by PAMBE in a chamber equipped with standard effusion cells for Ga and In, and a radio-frequency nitrogen plasma cell. Substrates consisted of 4-μm-thick GaN-on-sapphire templates. During the InGaN growth, the active nitrogen flux was fixed at $\Phi_N = 0.38$ ML/s, and the InGaN growth temperature was in the 610–670 °C range. The growth was monitored *in situ* by reflection high-energy electron diffraction (RHEED).

*Ex situ* measurements of the epilayer thickness and In alloy composition were performed by Rutherford backscattering spectrometry (RBS) using 2 MeV He+ ions and two detectors: a silicon *p-i-n* diode at a scattering angle of 165º, and a Si surface barrier detector at a scattering angle of 140º. Spectra were analyzed using the NDF code [15].

The surface morphology of the layers was studied by field emission scanning electron microscopy (FESEM) using a Zeiss Ultra 55 microscope, and by atomic force microscopy (AFM) using a Dimension 3100 system operated in the tapping mode. Structural properties were further analyzed by x-ray diffraction (XRD) using a SEIFERT XRD 3003 PTS-HR diffractometer with a beam concentrator prior to the Ge(220) two-bounce monochromator and a 0.15º Ge(220) long plate collimator in front of the detector.

For transmission electron microscopy (TEM), cross-section samples were thinned down to less than 10 µm using a Multiprep$^{TM}$ polisher. The electron transparency was achieved by ion milling at 5 kV using the GATAN precision ion milling system at an incidence angle of 5° with the sample holder, kept at 150 K in order to minimize the beam damage. A final step at low voltage (0.7 keV) and room temperature was applied in an attempt to further decrease the residual amorphous areas generated during the ion milling process. Observations were carried out by conventional transmission electron microscopy in a JEOL 2010 system. This work was complemented by analytical atomic resolution using high-angle annular dark field and energy-dispersive X-ray spectroscopy (EDS) in a JEOL ARM 200 scanning transmission electron microscope. Both microscopes were operated at 200 keV.

Optical transmission spectra were obtained at room temperature exciting with a 1000 W halogen lamp coupled with an Omni Lambda 300 monochromator. Photoluminescence (PL) measurements were performed exciting with a frequency-doubled continuous-wave Ar+ laser (wavelength $\lambda = 244$ nm) focused onto a 50 µm diameter spot. The PL signal was collected into a Jobin-Yvon monochromator with 45-cm focal length and detected by a charge-coupled-device camera.

## III. RESULTS AND DISCUSSION

### A. Growth parameters

The In composition in InGaN layers depends critically on the growth temperature [6,16-20], being therefore necessary to establish a reference parameter that guarantees its reproducibility. For the PAMBE growth of GaN, the substrate temperature in the range of 700−750°C is calibrated by the Ga desorption time measured by RHEED [21,22]. However, for the incorporation of In the substrate temperature should be reduced below 670°C, i.e. below the temperature where the Ga desorption rate is significant. In this case, a temperature reference can be provided by the In desorption time from the GaN(0001) surface.

To establish a calibration of the substrate temperature using the In desorption time, we performed a first experiment at a fixed substrate temperature ($T_S$ = 670 °C) exposing the GaN(0001) surface to In during a certain time ($t_{exp}$ = 1 min) long enough to form a stable In coverage on the surface, so that the desorption flux is equal to the impinging flux [23]. Note that the radio-frequency plasma cell was off during these experiments in order to prevent parasitic effects due to residual nitrogen. After shuttering the In cell, we record the transient in the RHEED specular beam intensity that reflects the surface evolution under vacuum during the In desorption, with the result depicted in Fig. 1(a). The duration of the desorption transient is related to the thickness of the original In layer. In the diagram in Fig. 1(a), we can observe

two stable regimes where small changes of the In flux have no effect on the RHEED desorption

transient [dashed lines in Fig. 1(a) marking the $\Phi_{In} = 0.044-0.080$ ML/s and the

$\Phi_{In} = 0.12-0.25$ ML/s windows]. From the length of the desorption time in the above-described

stability windows in Fig. 1(a) we can deduce the In coverage as a function of the In flux for a given temperature, following the method described in Refs. [23,24]: the $\Phi_{In} = 0.044-0.80$ ML/s range corresponds to an adsorbed In layer thickness of 1 ML, the $\Phi_{In} = 0.12-0.25$ ML/s range corresponds to an In coverage of 2 ML. Note that, within the 2 ML window, the transient variation of the RHEED intensity shows a characteristic shape, with a sharp local minimum associated to the desorption of the second In monolayer [outlined with squares in Fig. 1(b)],

followed by a larger and smoother minimum corresponding to the desorption of the first In

monolayer. Thus, within this window, the substrate temperature can be precisely determined

by measuring the time interval from the cell shuttering to the first sharp local minimum in the

RHEED intensity, independently of the impinging In flux, as illustrated in Fig. 1(b). To give an

idea of the precision of the method, Fig. 1(c) presents the time of the first inflexion point of the

In desorption transient as a function of the substrate temperature measured with a pyrometer.

For fixed N and Ga fluxes and working under In excess ($\Phi_{In} + \Phi_{Ga} > \Phi_N$), the In mole fraction of the layer is given by $x = 1 - \Phi_{Ga}/\Phi_N$ for temperatures below the InGaN decomposition temperature, which depends on the In mole fraction of the alloy. Taking into account our temperature calibration method, we have identified as the dashed line in Fig. 2(a) the growth temperature limit, $T_{lim}$, for a given In composition measured in thick (at least 250 nm) layers [25]. Note that at $T_{lim}$ the InGaN decomposition process is active.

As the substrate temperature decreases, the In desorption decreases, and there is risk of In droplet accumulation at the growth front [Fig. 2(b)]. On the other hand, for temperatures higher than $T_{lim}$, the In excess is not sufficient to compensate the rapid InGaN decomposition and is desorbed. Thus, the In mole fraction decreases and the growth rate decreases accordingly. Indications of InGaN decomposition are observed at the surface of the samples in Fig. 2(d). Best surface morphology [Fig. 2(c)] is achieved when the growth is performed close to $T_{lim}$, with a slight In excess corresponding to 1−2 ML of In adsorbed at the growth front, since this coverage reduces the surface free energy and enhances the adatom mobility [26,6].

Under these conditions, a series of $In_xGa_{1-x}N$ layers (series A) showing low-temperature PL emission in the 450−670 nm spectral range (corresponding to approximately $x = 0.13$–0.48) were synthesized, as displayed in Fig. 2(e). For the growth of these samples, the Ga flux was fixed to $\Phi_{Ga} = (1-x)\,\Phi_N$, and $\Phi_{In}$ was tuned to ensure metal excess and obtain smooth surfaces without In droplets or holes. The growth temperature was varied to approach the InGaN incorporation limit ($T_{lim}$).

Higher In compositions within the range under study result in a decrease of the low-temperature PL emission intensity by a factor of 20, and an increase of the PL full width at half maximum (FWHM) from 100 meV to 200 meV, probably due to In content inhomogeneities. This evolution of the optical properties is associated to an increase of the root-mean-square (rms) surface roughness from 1 to 5 nm (measured by AFM in a surface of $2\times2$ $\mu m^2$) for InGaN layers with a thickness ranging from ~150 nm to 250 nm.

In the following, this work focuses on InGaN layers containing $x \sim 0.30$ of In, the adequate mole fraction for the integration of InGaN in a tandem solar cell with silicon [2]. Therefore, for the studies below, the Ga flux was fixed at $\Phi_{Ga} = 0.7$ $\Phi_N$, whereas $\Phi_{In}$ was tuned to ensure metal excess (i.e. $\Phi_{In} >> 0.3$ $\Phi_N$). Unless indicated, the substrate temperature was kept close to $T_{lim} \sim 638°C$, the incorporation limit temperature for this alloy composition, as indicated in Fig. 1.

## B. Structural characterization and In incorporation

In a first set of experiments, the evolution of the InGaN in-plane lattice parameter ($a$) during the growth on GaN was determined by RHEED measuring the distance between the (10−10) and (−1010) streaks in the <11−20> azimuth, as illustrated in Fig. 3(a). The typical accuracy of this technique is of the order of $\Delta a/a_0 \sim 5\times10^{-4}$. Experimental results are presented in Fig. 3(b). Two regimes are identified: (1) an abrupt increase of $a$ by ~0.4% at a thickness of about 5 nm, which then remains stable during ~40 nm, followed by (2) a gradual increase of $a$ towards an almost relaxed steady-state situation reached at ~350 nm.

The evolution of the lattice parameter observed by RHEED can be related either to a gradual increase of the In incorporation in the layer, or to the relaxation of the strain generated by the lattice mismatch with the GaN substrate. In order to discern between composition and strain, a set of InGaN samples with different thicknesses (series B corresponding to samples S1-S4 in Table I: 30, 80, 120, and 500 nm thick, respectively) were grown on GaN-on-sapphire templates.

The In incorporation in the layers was assessed by RBS. Figure 4 displays the RBS spectra from $In_xGa_{1-x}N$ samples with thicknesses of 120 and 500 nm (S3 and S4, respectively), both showing increasing indium concentration towards the surface. In the case of S3, the indium profile is well modelled assuming that the sample consists of two InGaN layers (69 nm of $In_{0.242}Ga_{0.758}N$ followed by 36 nm of $In_{0.283}Ga_{0.717}N$). However, to reproduce the RBS profile of S4, it is necessary to assume four $In_xGa_{1-x}N$ layers with In mole fractions evolving from x = 0.214 to 0.315 (121 nm of $In_{0.211}Ga_{0.789}N$ followed by 75 nm of $In_{0.244}Ga_{0.756}N$, 145 nm of $In_{0.285}Ga_{0.715}N$, and 104 nm of $In_{0.313}Ga_{0.687}N$). The average In composition in the layers and the minimum/maximum compositions assumed in the multilayer modelling are summarized in Table I. Converting the measurements of S4 in Fig. 4(b) into discrete variations of the in-plane lattice parameter assuming complete relaxation (using $a_{GaN} = 3.189$ Å and $a_{InN} = 3.545$ Å [27]), we obtain the solid black line in Fig. 3(b). The deviation of the RHEED experimental measurements from such line, depicted as a shadowed area in Fig. 3(b), reflects the InGaN elastic strain. We can therefore conclude that the mismatch relaxation process takes place gradually along the first ~350 nm of the layer. This is in contrast with the case of InN grown on GaN, where most of the in-plane lattice mismatch is immediately relaxed by a network of misfit dislocations at the interface [28].

To get further information about the *ex-situ* strain state, the layers have been analyzed by XRD. Measurements of the symmetric (000$l$) reflections reveal that both GaN and InGaN lattices are significantly tilted with respect to the substrate Al$_2$O$_3$ surface plane. The GaN-on-sapphire layers present a tilt of 0.66±0.03° towards the [10−10] direction, and the InGaN layers grown on top tend to tilt towards [−1010], partially correcting the initial GaN tilt. Table I presents a summary of the relative tilt between the InGaN and the GaN lattices, which increases from being negligible for 30 nm thick InGaN to a value of 0.47° for the 500 nm thick layer (sample S4).

Reciprocal space maps have been recorded around the (−1015) asymmetric reflection, as illustrated in Fig. 5(a-d). For an appropriate interpretation of the data, the GaN and InGaN reflections have been corrected to account for the tilt of their respective lattices. Superimposed straight vertical and diagonal lines correspond to the position of the InGaN diffraction points assuming that the InGaN layer is fully strained on GaN or completely relaxed, respectively, for In mole fractions ranging from 5 to 30%. Dashed lines represent the evolution from the state strained on GaN to completely relaxed for a given In composition. The vertical alignment (same $Q_x$ projection of the reciprocal space vector) of the InGaN and the GaN reflections displayed in Fig. 5(a) reveals that the first 30 nm of InGaN are pseudomorphically grown on GaN (sample S1). However, reciprocal space maps 5(b), (c) and (d) show an increasing relaxation of the lattice, which shifts the InGaN reflection towards lower $Q_x$ values, together with an increase of the average In content. For the thicker layer (S4), XRD measurements point to a possible In demixing phenomena along the growth axis.

The average In mole fraction in the layers and their average strain state can be extracted from the $a$ and $c$ lattice parameters obtained from the reciprocal space maps assuming biaxial strain ($e_{zz}/e_{xx} = -2c_{13}/c_{33}$, with $e_{zz}$ and $e_{xx}$ being the strain along <0001> and <11−20>, and $c_{13}$ and $c_{33}$ being the elastic constants of the InGaN film). Assuming the lattice parameters along the growth axis $c_{GaN} = 5.185$ Å and $c_{InN} = 5.703$ Å [27]; the elastic constants $c_{13} = 106$ and $c_{33} = 398$ for GaN [27], and $c_{13} = 91$ and $c_{33} = 243$ for InN [29]; and linear interpolations for ternary alloys, the extracted values of In mole fraction (minimum, average, and maximum), and average relaxation [30] are summarized in Table I. The obtained XRD results show good agreement with RBS measurements.

The thickness of the InGaN samples measured by XRD (S1-S2) and RBS (S3-S4) (listed in Table I) is systematically smaller than the nominal thickness, which is an indication of the InGaN decomposition, more accentuated in thinner samples.

In summary, a compositional gradient associated to strain is observed in InGaN layers,

similar to the compositional pulling effect reported in GaAs- and InP-based material systems [31,32]. The first 40 nm are almost pseudomorphic on GaN and homogeneous in In composition, and the progressive misfit relaxation between 40 nm and 350 nm is results in an enhancement of In incorporation.

The relaxation process is associated with an increase of the mosaicity of the layers with the thickness, as revealed by the reciprocal space mapping of Fig. 5, and an increase of the rms surface roughness from 0.5 to 3 nm, as represented in Fig. 6(a). The surface morphology of samples S1 and S4 is displayed in Figs. 6(b) and (c), respectively.

In order to investigate the generation of defects associated to the relaxation process, the samples were analyzed by TEM. In contrast to other techniques, TEM only exhibits the local state of the observed areas. The growth of InGaN alloys is generally associated to structural defects such as misfit and threading dislocations [33], stacking faults which may form in the basal [34] or prismatic planes [35], but also to ordering [36] and/or phase separation [37]. Sample S1 (30 nm thick) shows basal stacking defects close to the interface; they appear to be mostly mediated by threading dislocations from the GaN buffer layer underneath, as it can be observed in the $(1-100)$ weak beam micrograph in Fig. 7(a). This probably means that the initial relaxation observed by RHEED in the first few nanometers of growth is associated to misfit dislocations confined at the InGaN/GaN interface. At 80 nm (sample S2), the interfacial area in Fig. 7(b) is marked by the appearance of misfit dislocations, basal stacking defects which may spread up to the surface, and additional defects form inside the layer not directly connected to the interface. Therefore, the plastic relaxation of these layers through misfit dislocations takes place at a thickness between 30 and 80 nm in agreement with the above observations.

This type of relaxation is amplified at larger thicknesses, as it can be seen in Fig. 8 for 200 nm thick InGaN layers. The (0002) weak beam micrograph in Fig. 8(a) shows numerous threading dislocations with a screw component, which are generated inside the InGaN layer and propagate towards the surface. Furthermore, the (10−10) weak beam micrograph in Fig. 8(b) shows basal stacking defects together with threading dislocations with an edge component, which contribute to the progressive strain relaxation. In the case of S4 (500 nm thick InGaN), the (10−10) weak beam micrograph in Fig. 8(c) shows a high density of basal stacking defects which are evenly distributed in the layer, and present a non-uniform but higher density towards the surface. These stacking defects are not usual basal stacking faults [38]; they constitute a random disruption of the hexagonal-close-packed stacking sequence along the growth axis. In this image, they are connected with the visible threading dislocations which have an edge component. Moreover, threading dislocations originate mostly at the interface in relation with the misfit dislocations, and this complex defect system probably contributes to the progressive strain relaxation, as pointed out by RHEED and XRD.

Along with this defect analysis, we also have carried out EDS measurements of the local In composition by recording profiles from the interface towards the surface of the layers. As indicated in Table I, there are large error bars and a certain disagreement with larger-scale techniques, pointing to large In composition fluctuations inside the InGaN layer, which may in turn explain the formation of the stacking defects.

## C. Optical characterization

Room-temperature optical transmission and PL measurements of the InGaN samples under study are presented in Fig. 9. Transmission spectra show the sharp absorption cutoff of GaN at 365 nm, followed by the absorption associated to InGaN for wavelengths $\lambda < 450$ nm. The average band gap energy ($E_g$) and absorption edge broadening ($\Delta E$, equivalent to the Urbach tail energy) of the InGaN layers were estimated from a sigmoidal approximation of the absorption spectra [39], with the results presented in Table II. The shift of the band gap from 2.62 to 2.4 eV for increasing the layer thickness is consistent with the evolution of the average In mole fraction and strain state. On the other hand, $\Delta E$ increases with the sample thickness as expected from the higher In inhomogeneity in thicker samples.

The room-temperature PL spectra of the InGaN layers, displayed in Fig. 9(b), presents a red shift of the main peak for increasing thickness, confirming the enhancement of the In incorporation beyond the first 80 nm. Comparing the PL peak wavelength at room temperature ($\lambda_{PL}$ in Table II) with the values of $E_g$ extracted from transmission measurements, we observe an increase of the room-temperature Stokes shift with the layer thickness from 120 meV (S2) to 270 meV (S4). The obtained values are slightly lower than those reported by Martin *et al.* for InGaN layers emitting at similar energies [40].

The optical properties were further investigated by temperature-dependent PL measurements. The low-temperature (T = 7 K) PL spectra of the InGaN samples is shown in the inset of Fig. 10(a), whereas the evolution of the normalized integrated PL intensity with the temperature is presented in Fig. 10(a). For the analyzed samples, the thermal extinction of the PL intensity, *I(T)*, can be reproduced considering two nonradiative recombination channels [18]:

$$I(T) = \frac{I(T=0K)}{1 + ae^{-E_{loc1}/k_BT} + be^{-E_{loc2}/k_BT}} \tag{1}$$

where $E_{loc1}$ and $E_{loc2}$ represent the average energetic barriers that the carriers must surmount in order to escape from its localization and reach the nonradiative recombination centers, $k_BT$ is the thermal energy, and *a* and *b* are fitting constants. The agreement of experimental data to Eq. (1) is illustrated by the solid lines in Fig. 10(a). Values of $E_{loc1}$ = 10±5 meV (for T<100K) and $E_{loc2}$ = 124±10 meV (for T>100K) are obtained for the samples under study, as summarized in Table II. The small localization energy is in the same order of magnitude than the obtained in bulk InN films grown by PAMBE [41].

The evolution of the emission peak energy as a function of temperature describes an S shape where the blue shift at intermediate temperatures is explained by the filling of potential valleys with different depths upon excitation [42,43]. For the quantification of these potential fluctuations, Eliseev *et al.* [42] proposed a band-tail model assuming a Gaussian-like distribution of the density of states. This results in a correction to Varshni's equation by $-\sigma^2/kT$, where $\sigma$ is the dispersion of the Gaussian band-tail density of states. From the analysis of InGaN layers with the different thicknesses, an average value of $\sigma = 100\pm20$ meV is obtained, *i.e.* in the same order of magnitude as the above estimated localization energy $E_{loc2}$.

## IV. CONCLUSIONS

The interplay between In incorporation and strain relaxation kinetics in high-In-content InGaN layers grown by plasma-assisted molecular-beam epitaxy on GaN-on-sapphire is reported. For In mole fractions $x = 0.13$–$0.48$, best structural and morphological quality is obtained under In excess conditions, and at a growth temperature where InGaN decomposition is active. Under these growth conditions, both *in situ* and the *ex situ* experiments point to an onset of strain relaxation at 40 nm, which stabilizes at an almost relaxed value after 350 nm. This relaxation is associated to the appearance of misfit dislocations, stacking faults, edge- and screw-type threading dislocations, a tilt of the lattice towards [1−100], mosaicity and surface roughening (rms increases from 1 to 5 nm). Moreover, the inhomogeneous strain state along the growth axis perturbs the alloy composition, which evolves from $x = 0.22$ to 0.32. In atoms are excluded from the InGaN film under compressive stress to minimize the lattice mismatch between the epitaxial layer and the substrate. So the layer only reaches its nominal composition when is almost fully relaxed. Optically, the composition gradient translates into a red shift of the emission/absorption at room temperature ($E_g = 2.62-2.4$ eV, $\Delta E = 54-123$ meV, $\lambda_{PL} = 498-551$ nm), and an increase of the Stokes shift (120−270 meV) as a function of the layer thickness (30−500 nm). This information is particularly relevant for the synthesis of high-In-content quantum wells or thin films in the active region of emitters and detectors, since their thickness and composition depend not only on the impinging fluxes and growth temperature, but also on the strain state imposed by the underlayers.

## ACKNOWLEDGEMENTS

The authors would like to acknowledge Y. Cure for useful discussions and technical support on the RHEED analysis. Partial financial support was provided by the DSM-Energie "NANIPHO" project, by the Marie Curie IEF grant "SolarIn" (#331745), and by the French National Research Agency via the GANEX program (ANR-11-LABX-0014). KL and EA acknowledge support by FCT Portugal (grants PTDC/FIS-NAN/0973/2012 and Investigador FCT).

20% in thick layers. Furthermore, precise decomposition temperature must be taken with caution, since it can depend on the overall metal-to-nitrogen ratio and on the growth rate, as described in Ref. 20.

**TABLES**

**Table I:** Summary of structural data of In$_x$Ga$_{1-x}$N samples with different thicknesses: Nominal and measured thickness (measured by XRD for S1 and S2 and by RBS for S3 and S4), In mole fraction measured by RBS (minimum, average, maximum), tilt towards [−1100] direction respect to the GaN underlayer, average relaxation and In mole fraction measured by XRD (minimum, average, maximum) and In mole fraction measured by TEM (minimum, average, maximum). Average values of In mole fraction obtained by XRD correspond to the more intense diffraction peak.

| Sample | Nominal (measured) Thickness (nm) | RBS | | | XRD | | | | | EDS-TEM | | |
|---|---|---|---|---|---|---|---|---|---|---|---|---|
| | | $x_{min}$ | $\bar{x}$ | $x_{max}$ | Tilt | Average Relaxation (%) | $x_{min}$ | $\bar{x}$ | $x_{max}$ | $x_{min}$ | $\bar{x}$ | $x_{max}$ |
| S1 | 30 (25) | - | - | - | 0° | 0±0.7 | 0.214 | 0.224 | 0.234 | 0.12 | 0.23 | 0.36 |
| S2 | 80 (70) | - | - | - | 0.05° | 24±5 | 0.216 | 0.219 | 0.241 | 0.14 | 0.24 | 0.32 |
| S3 | 120 (105) | 0.242 | 0.256 | 0.283 | 0.41° | 48±8 | 0.224 | 0.250 | 0.297 | 0.16 | 0.24 | 0.33 |
| S4 | 500 (445) | 0.228 | 0.269 | 0.324 | 0.47° | 52±9 | 0.180 | 0.268 | 0.320 | 0.16 | 0.22 | 0.28 |

**Table II:** Optical parameters extracted from the InGaN samples under study: average band gap ($E_g$), absorption edge broadening ($\Delta E$), PL peak wavelength at room temperature ($\lambda_{PL}$), activation energies of the nonradiative processes ($E_{loc1}$, $E_{loc2}$), and localization energy extracted for Eliseev's equation ($\sigma$).

| Sample | Nominal Thickness (nm) | $E_g$ (eV) | $\Delta E$ (meV) | $\lambda_{PL}$(nm) | $E_{loc1}$ (eV) | $E_{loc2}$ (meV) | $\sigma$ (meV) |
|--------|------------------------|------------|------------------|--------------------|-----------------|------------------|----------------|
| S1 | 30 | 2.62±0.01 | 54±4 | - | - | - | - |
| S2 | 80 | 2.61±0.05 | 99±15 | 498±5 | 8±5 | 125±10 | 120±20 |
| S3 | 120 | 2.55±0.07 | 91±7 | 523±5 | 6±5 | 123±10 | 90±20 |
| S4 | 500 | 2.4±0.1 | 123±10 | 551±5 | 15±5 | 125±10 | 95±20 |

**FIGURE CAPTIONS**

**Figure 1:** (a) Evolution of the RHEED intensity during In desorption from a GaN(0001) surface after 1 min exposure to the In flux indicated in the right column measured at a substrate temperature of $T_S = 670$ °C. The right column indicates the In flux during In exposure. Dashed lines are eye guides to highlight the two stable regions. (b) Evolution of the RHEED intensity during the In desorption after 1 min exposure to an In flux of $\Phi_{In} = 0.38$ ML/s, measured at the different substrate temperatures indicated in the right column. Open squares are eye guides to highlight the first inflexion point of the In desorption transient. In both cases the In shutter is closed at $t = 0$ s. (c) Time of the first inflexion point of the In desorption transient [open square in (b)] as a function of the substrate temperature.

**Figure 2**: (a) Influence of the growth temperature on the In incorporation under In-rich conditions. The dashed line indicates the maximum In incorporation as a function of the temperature. The colored squares correspond to the colored PL lines in Fig. 2(e). For a given In content, when increasing the temperature the InGaN surface evolves from (b) In accumulation to (c) smooth, and (d) holes due to InGaN decomposition. (e) Normalized low-temperature PL emission of $In_xGa_{1-x}N$ films ($x$ evolving from 0.13 to 0.48), covering the whole visible spectrum. Top: photograph of the samples.

**Figure 3**: (a) Schematic of the method used to extract the in-plane lattice parameter variation, $\Delta a/a$, from the RHEED pattern. The distance between outer streaks, proportional to $1/a$, is determined by fitting the intensity profile with Pseudo-Voigt curves. (b) Variation of the in-plane lattice parameter of the InGaN layer deposited on GaN as a function of the thickness obtained by in situ RHEED measurements. The straight solid lines are the result of converting the In composition extracted from RBS measurements of S4 into variations of the in-plane lattice parameter assuming complete relaxation (using $a_{GaN} = 3.189$ Å and $a_{InN} = 3.545$ Å [22]).

**Figure 4:** Measured (dots) and simulated (solid line) Rutherford backscattering spectra of (a) 120 nm and (b) 500 nm thick InGaN layers (S3 and S4 in Table I). Experimental data were simultaneously measured with two detectors: a silicon *p-i-n* diode at a scattering angle of 165°, and a Si surface barrier detector at a scattering angle of 140°. Solid lines are simultaneous fits of the spectra using the NDF code [15]. The arrows in (a) represent the Ga and In surface channels. Dashed and dotted lines in (b) represent the contribution from Ga and In atoms to the spectra, respectively.

**Figure 5:** Reciprocal space map around the asymmetric (−1015) reflection for InGaN layers with different thicknesses: (a) 30, (b) 80, (c) 120 and (d) 500 nm (samples S1–S4 in Table I, respectively). $Q_x$ and $Q_z$ are the reciprocal space vectors in Å$^{-1}$. Solid straight and diagonal lines correspond to the position of the InGaN diffraction points assuming that InGaN is strained or relaxed on GaN, respectively. Dashed lines represent the evolution of the strain state for 5%, 10%, 15%, 20%, 25% and 30% In content in InGaN.

**Figure 6:** (a) Evolution of the rms surface roughness with the layer thickness. AFM images of an area of $10 \times 10$ µm² of the (b) 30 nm thick and (c) 500 nm thick InGaN layers (samples S1 and S4 in Table I, respectively).

**Figure 7:** Weak beam TEM micrographs with g = 10−10 for (a) a 30 nm thick InGaN layer (S1) exhibiting stacking defects at the InGaN/GaN interface, (b) a 80 nm thick InGaN layer (S2) showing stacking defects (S), and misfit (M) and threading (D) dislocations. The white arrows underline the GaN/InGaN interface.

**Figure 8:** Weak beam TEM micrographs of a 200 nm thick InGaN layer with (a) g=0002, showing numerous threading dislocations with a screw component, and (b) g=10−10, showing the stacking defects which have formed all over the layer and interact with edge component threading dislocations. (c) Weak beam micrograph with g=10−10 of a 500 nm thick InGaN layer (S4) showing the multiplication of edge component threading dislocations (D) which originate from the interface, they are connected to misfit dislocations (M) and interact with the stacking defects (S) whose density increased dramatically. The white arrows underline the GaN/InGaN interface.

**Figure 9:** Room-temperature transmittance (a) and photoluminescence (b) spectra of the InGaN structures (samples S1−S4 described in Table I). The room-temperature PL emission of sample

S1 is not measurable by our PL setup. The oscillations superimposed to the PL spectra are due to the Fabry-Perot interference associated to the nitride epilayer thickness.

**Figure 10:** (a) Thermal evolution of the integrated PL emission of the analyzed InGaN structures (samples S1−S4 described in Table I). Solid lines are fits to Eq. (1). Inset: Low temperature ($T$ = 7 K) PL spectra of the samples. (b) Temperature dependence of the PL emission peak energy. Dashed lines indicate the fits to Eliseev's correction [42] to model the InGaN alloy fluctuations.

**Figure 1**

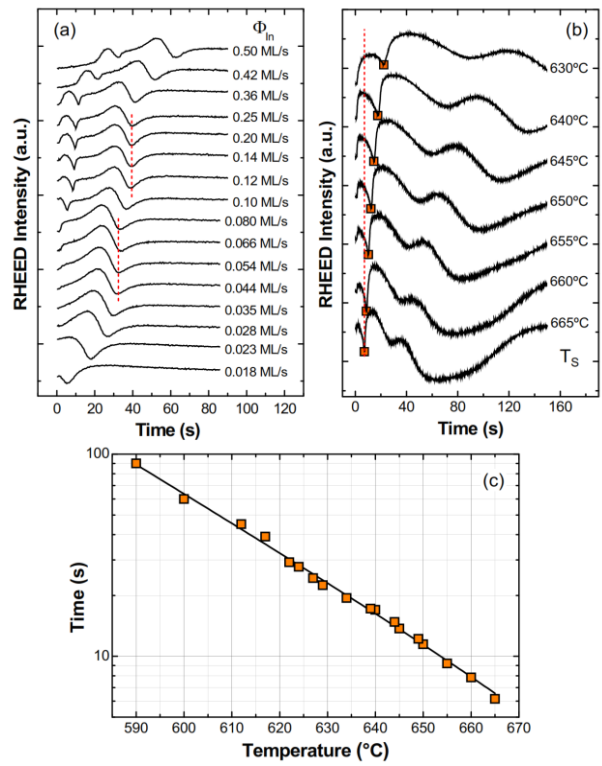

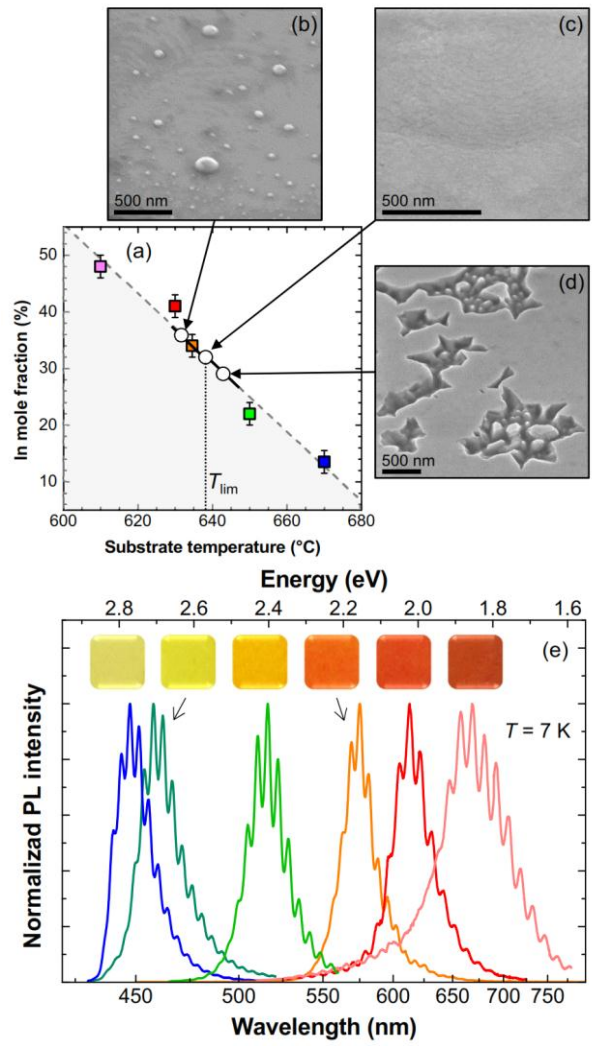

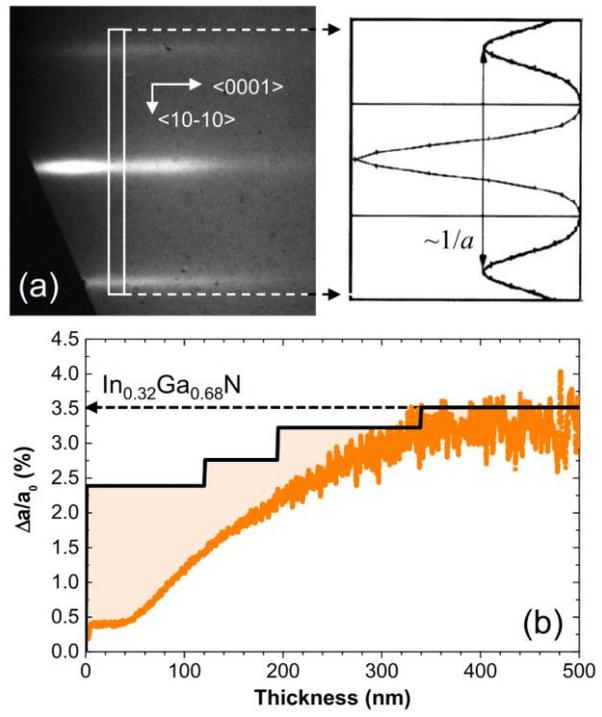

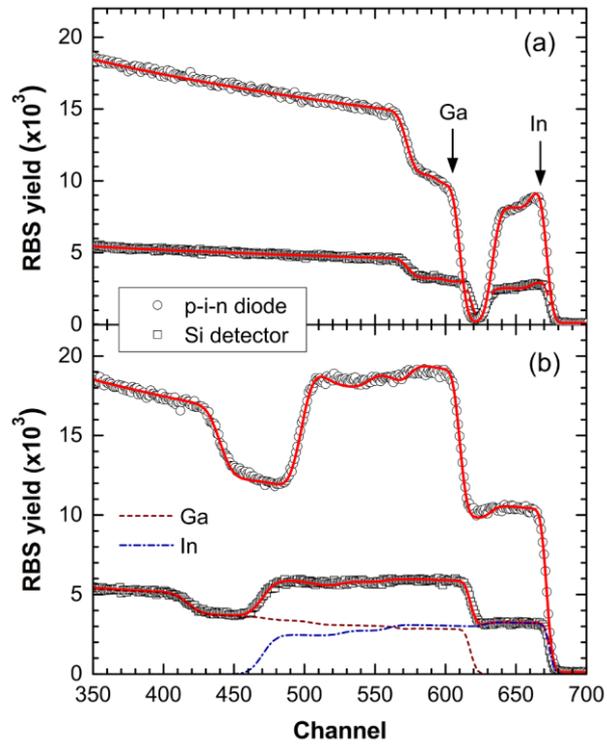

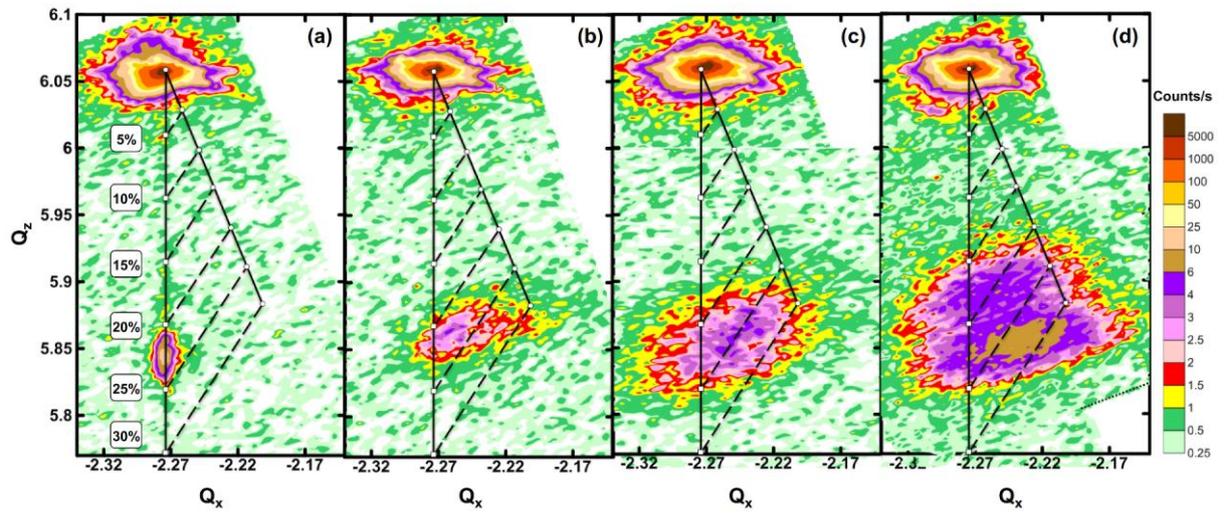

**Figure 6**

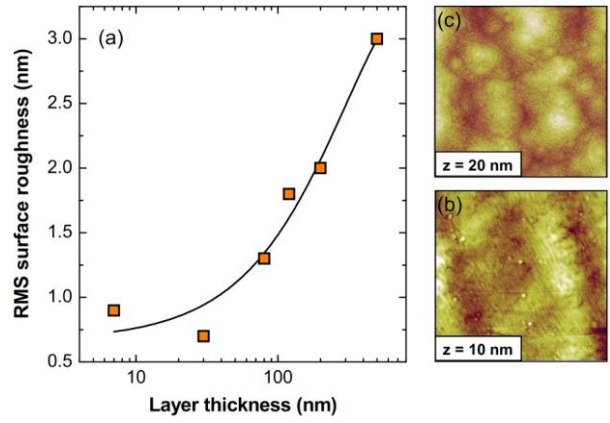

**Figure 7**

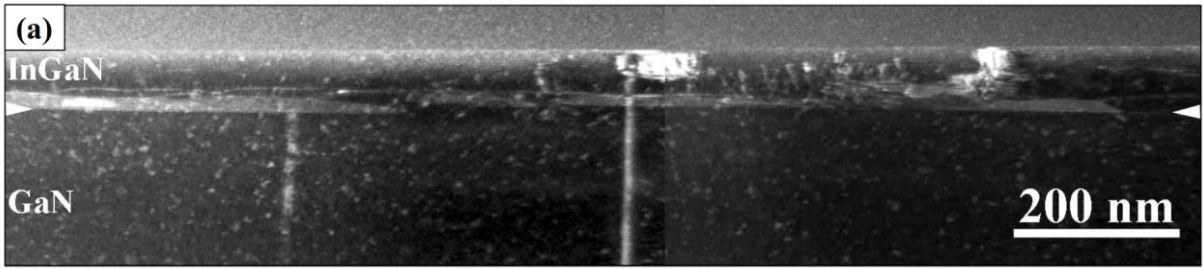

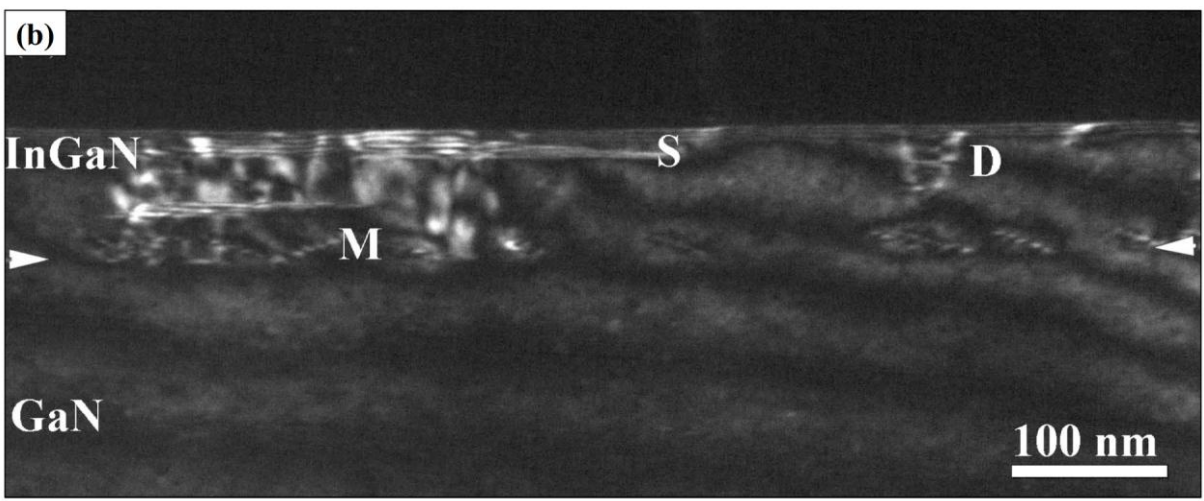

**Figure 8**

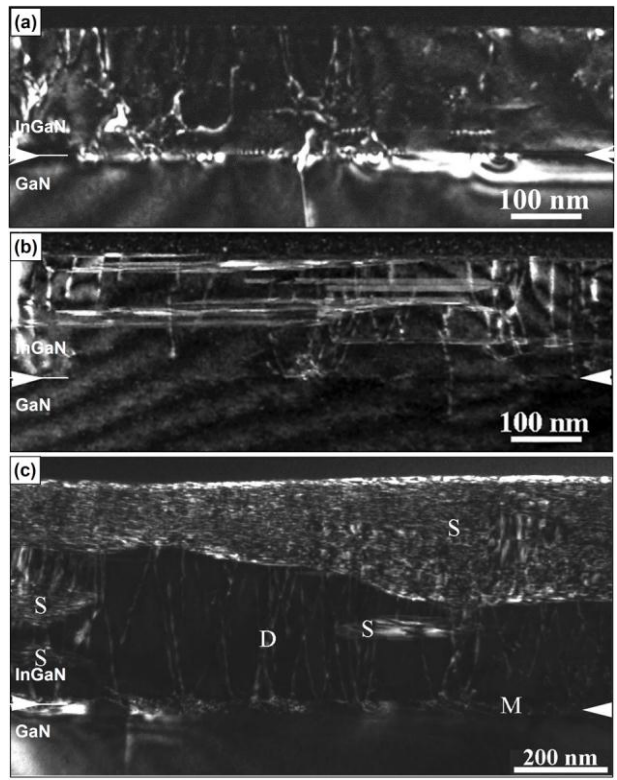

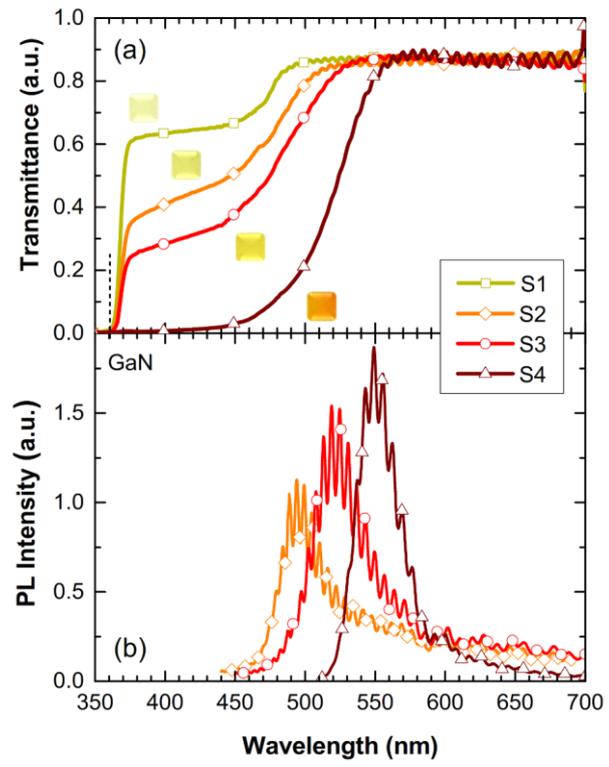

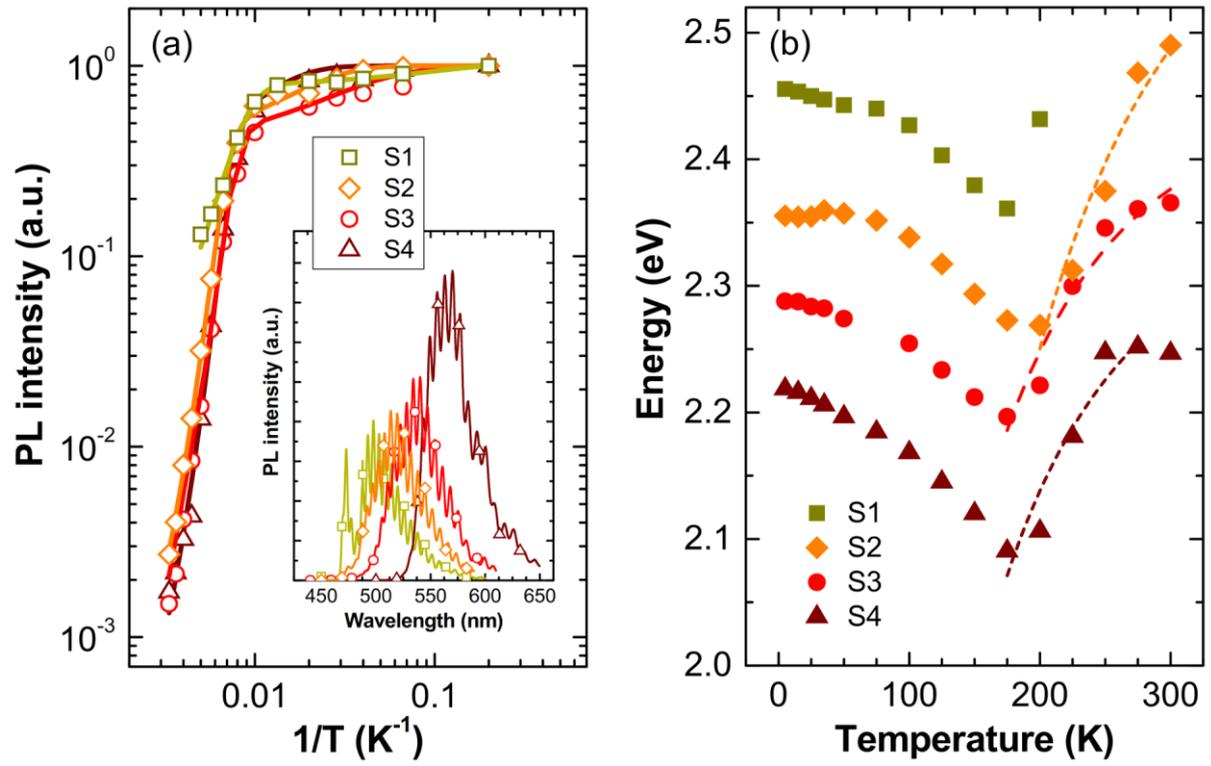